\documentclass{article}
\usepackage{romp}
\usepackage{amsmath, amsgen,amstext,amsbsy,amsopn,amsthm, amssymb}

\newcommand{\eps}{\epsilon}

\newcommand{\aoo}{{a_{\bf 0}^{\phantom *}}}
\newcommand{\aos}{a_{\bf 0}^*}
\newcommand{\ak}{{a_{\bf k}^{\phantom *}}}
\newcommand{\aks}{a_{\bf k}^*}
\newcommand{\aq}{a_{\bf q}}

\newcommand{\no}{n_{\bf 0}}
\newcommand{\kk}{{\bf k}}
\newcommand{\pp}{{\bf p}}
\newcommand{\qq}{{\bf q}}
\newcommand{\hi}{\mathcal{H}}
\newcommand{\hip}{{\mathcal{H}'}}

\newcommand{\half}{\mbox{$\frac{1}{2}$}}

\newcommand{\x}{{\bf r}}

\newcommand{\e}{\widetilde{e}}

\newcommand{\Tr}{{\rm Tr} }

\title{Bose-Einstein Condensation \\ and Spontaneous Symmetry Breaking$^{1}$}
\author{ Elliott H.~Lieb\thanks{ \hskip0.1cm Work partially supported by U.S. National Science Foundation grant PHY 01 3998}\\ Department of Physics,
Jadwin Hall, Princeton University,\\
P.~O.~Box 708, Princeton, New Jersey 08544 \\ e-mail: lieb@princeton.edu
 \\[2ex]
         Robert Seiringer\thanks{ \hskip0.1cm Work partially supported by U.S. National Science Foundation grant PHY 03 53181, and by
an A.P. Sloan Fellowship}\\Department of Physics,
Jadwin Hall, Princeton University,\\
P.~O.~Box 708, Princeton, New Jersey 08544\\ e-mail: rseiring@princeton.edu \\[2ex]
    Jakob Yngvason\thanks{ \hskip0.1cm Work partially supportd by  FWF grant
P17176-N02 and EU grant HPRN-CT-2002-00277}                  \\ Institut f\"ur Theoretische Physik,
Universit\"at
Wien, \\Boltzmanngasse 5, A-1090 Vienna, Austria\\ e-mail: yngvason@thor.thp.univie.ac.at }

\begin{document}

\maketitle

\renewcommand{\thefootnote}{${1}$}
\footnotetext{\scriptsize{Contribution to the proceedings of the 21st Max Born Symposium, Wroclaw, Poland, June 26--28, 2006. Talk given by J. Yngvason. \copyright\, 2006 by the authors. This paper may be reproduced, in its
entirety, for non-commercial purposes. }}

\begin{abstract}
After recalling briefly the connection between spontaneous symmetry breaking and off-diagonal long range order for models of magnets a general proof  of  spontaneous breaking of gauge symmetry as a consequence of Bose-Einstein Condensation is presented. The proof is based on a rigorous validation of Bogoliubov's $c$-number substitution for the $\kk={\bf 0}$
mode operator $\aoo$.
\end{abstract}

\section{Introduction}

The connection between  Bose-Einstein Condensation (BEC) and spontaneous gauge symmetry breaking (GSB) is often taken for granted, usually with the following argument that appeals  to Bogoliubov's pioneering work \cite{bog}: BEC means that in the thermodynamic limit the density $V^{-1}\langle\aos \aoo\rangle$ of particles in the mode of lowest momentum $\kk={\bf 0}$ is different from zero. The spontaneous breaking of gauge symmetry, on the other hand, is associated with the non-vanishing of $V^{-1}|\langle\aoo \rangle|^2$ in the thermodynamic limit. Here it is important that the expectation value is  understood in the sense of Bogoliubov's \lq quasi averages\rq, i.e., with a symmetry breaking field that is taken to zero {\it after}  the thermodynamic limit.  By the Cauchy-Schwarz inequality, we always have $|\langle\aoo \rangle|^2\leq \langle\aos \aoo\rangle$. Thus GSB implies BEC. On the other hand, if BEC holds, the  macroscopic occupation of the zero mode is claimed to justify the replacement of the operators $\aoo$ and  $\aos$ by c-numbers and hence equating  $|\langle\aoo \rangle|^2$ with $\langle\aos \aoo\rangle$. Therefore BEC and GSB are equivalent.

A rigorous justification of  of Bogoliubov's $c$-number substitution is not a trivial matter, however,  and the reasoning above is clearly not a mathematical proof. In a classic paper \cite{gin}, that is less known in the physics community than it deserves,   Ginibre established the validity of the substitution as far as  the computation of the pressure in the thermodynamic limit is concerned. As Ginibre pointed out, it is not necessary to assume macroscopic occupancy of the zero mode (although the substitution is only a useful computational tool if this is the case.) In a recent paper \cite{lsy} (see also \cite{book}, Appendix D) we presented a  simple and general proof of the exactness of $c$-number substitutions in bosonic Hamiltonians in the thermodynamic limit, using the Berezin-Lieb inequality \cite{BL,Li, simon} (which was only derived after Ginibre's paper appeared). We then proved the equivalence of GSB and BEC, where the expectation values are defined as quasi averages in the sense of Bogoliubov, and moreover  that BEC in the standard sense {\it without} a symmetry breaking field implies BEC in the quasi average sense and hence GSB. This last statement, although intuitively plausible \cite{roepstorff}, does not seem to have been rigorously verified before.  
In brief:
\begin{equation}\label{mainstatement}{\rm BEC}\rightarrow ({\rm BEC})_{\rm qa}\leftrightarrow{\rm GSB}.\end{equation}
The present  contribution to the Proceedings of the 21st Max Born Symposium gives  an account of  the proof. For comparison and for pedagogical reasons we start by recalling briefly the connection between spontaneous symmetry breaking and off-diagonal long range order for a model of a ferromagnet. An important difference between this case and that of a Bose gas is that the symmetry breaking term in the latter case does not commute with the Hamiltonian.

Another proof of  $({\rm BEC})_{\rm qa}\leftrightarrow{\rm GSB}$, based on Ginibre's original results about the $c$-number substitutions,  was given by S\"ut\H o in \cite{suto}. A rather different approach, with focus on operator algebraic aspects and correlation inequalities,  can be found in a 1982 paper by Fannes, Pul\'e and Verbeure \cite{fannes}, see also a recent generalization to nonhomogeneous systems  by Pul\'e, Verbeure and Zagrebnov \cite{puleetal}. In these papers the equilibrium states on the Weyl algebra for the infinite system and their decomposition into symmetry breaking states are studied.  Some technical assumptions about these states are required in \cite{fannes} and \cite{puleetal}  but in Remark 2.1 in \cite{puleetal} it is claimed that the conclusions are valid under weaker assumptions.  The algebraic approach sheds a different light on the problem and
is complementary to our method where essentially the only  prerequisites are the existence of  the thermodynamic limit of the grand canonical pressure and average density.

Finally we remark that the Berezin-Lieb inequality used in our approach gives explicit error bounds that allow  us to go beyond \cite{gin} and make $c$-number
substitutions for {\it many} $\kk$-modes at once, provided the number
of modes is lower order than the particle number. This method,  that was used in a different  context in \cite{LT}, has recently proved to be a powerful tool for solving other problems about the Bose gas \cite{LSeir, Seir}.

\section{Symmetry Breaking and Off-diagonal Long Range Order in Magnets}

The concept of {\it spontaneous symmetry breaking} means quite generally that a symmetry of the Hamiltonian or Lagrangian of a system is not 
present in the state under consideration (usually a ground state or a 
thermal equilibrium state).  A Heisenberg ferromagnet with a Hamiltonian of the  form
\begin{equation}H=H_0-BM \quad{\rm with}\quad  H_0=-\sum_{ x\neq y\in\Lambda}J_{xy}\vec S_x\cdot \vec S_y
\quad{\rm and}\quad 
M=\sum_{x\in\Lambda}S^{(3)}_x
\end{equation}
provides a well known example to illustrate this phenomenon.
Here $\Lambda\subset \mathbb Z^d$ is a lattice cube of finite size  $|\Lambda|$ and  periodic boundary conditions are assumed. The spin operators
$\vec S_x=(S^{(1)}_x,S^{(2)}_x,S^{(3)}_x)$ with $[S^{(1)},S^{(2)}]={\rm i} S^{(3)}$ etc act on a copy of ${\mathbb C}^{2s+1}$  (with $s\in\{\half,1,\dots\}$) at each lattice site and operators on different  sites commute. The coefficients $J_{x,y}\geq 0$ measure the interactions between different sites and $B$ denotes the strength of a magnetic field in the 3-direction. If $B=0$ the Hamiltonian 
is invariant under the transformations  $\vec S_x \to R \vec S$, 
$R\in O(3)$. The last term in the Hamiltonian explicitly breaks this symmetry if $B\neq 0$.

The thermal equilibrium state at temperature $T$ in the finite lattice is defined by
\begin{equation}\langle A\rangle_{B,T,\Lambda}=\frac {{\Tr\,} Ae^{-H/kT}}{{\Tr}\, e^{-H/kT}}\end{equation}
and the magnetization, i.e., the  thermal average of the magnetic moment per spin, is
\begin{equation}{} m_{\Lambda}(B,T)=|\Lambda|^{-1}\langle M
\rangle_{B,T,\Lambda}\, .\end{equation}
The Gibbs free energy per spin, 
 \begin{equation}g_\Lambda(B,T)=-|\Lambda|^{-1}kT\ln{\rm Tr}\,e^{-H/kT}\end{equation}
is a concave function of $B$ with 
\begin{equation}{} m_{\Lambda}(B,T)=-\frac{\partial g_\Lambda(B,T)}{\partial B}.\end{equation}
Since $T$ may be regarded as fixed in the following we shall drop it in the notation, and in order to avoid irrelevant factors, choose energy units so that  $kT=1$.

For a finite $\Lambda$, ${}{m}_{\Lambda}(B)$ is continuous in $B$, and for $B=0$ we always have
\begin{equation}{}{m}_{\Lambda}(0)=0\end{equation}
because the state is invariant under a unitarily implemented rotation that takes $S^{(3)}_x\to -S^{(3)}_x$. 
On the other hand, the thermodynamic limit of the magnetization
\begin{equation}{} m(B)=\lim_{\Lambda\to {\mathbb Z}^d}{} m_{\Lambda}(B)\end{equation}
may break the symmetry for sufficiently low $T$, i.e., 
it is possible that 
\begin{equation}{} m_{\pm} \equiv  \lim_{B\to 0^{\pm} }{} m(B)\neq 0.\end{equation}
Since the Hamiltonian is invariant under simultaneous flipping of the spins and $B$ we always have ${} m_{-}=-{} m_{+}$, so if ${} m_{\pm}\neq 0$ the  Gibbs free energy in the thermodynamic limit, $g(B)=\lim_{\Lambda\to {\mathbb Z}^d}g_\Lambda(B)$,  is not differentiable at $B=0$. As a concave function it is, however, differentiable almost everywhere.

We may also consider the average of the square of  the magnetic moment per spin, i.e., 
\begin{equation}{} {m^2_{\Lambda}}(B)\equiv\big\langle (M/|\Lambda|)^2 \big\rangle_{B,T,\Lambda}=\Big\langle |\Lambda|^{-2} 
\sum_{x,y\in\Lambda}S^3_xS^3_y
\Big\rangle_{B,T,\Lambda}\end{equation}
with a corresponding thermodynamic limit  $ {}{ m^2}(B)$. By the Cauchy-Schwarz inequality we always have
\begin{equation}{}{ m^2}(B)\geq \left({} m(B)\right)^2.\end{equation}
Hence, {\it spontaneous symmetry breaking} in the sense that $\lim_{B\to 0^+}{} m(B)\neq 0$, implies {\it off-diagonal long range order} in the sense that
\begin{equation}\lim_{B\to 0^+}{}{m^2}(B)>0.\end{equation}
It is important to note, however, that  it is {\it a priori} not clear that this implies
${}{m^2}(0)>0$ for reasons  explained further below.

On the other hand, a general argument due to Griffiths \cite{griffiths} shows that
\begin{equation}\label{grmagn}{}{m^2}(0)\leq \lim_{B\to 0^+}{}{m^2}(B) =\lim_{B\to 
0^+}\left({} m(B)\right)^2.\end{equation}
The argument is based on two facts. The first  is the commutativity of the symmetry breaking term, $BM$, with  the other part, $H_0$,  of the Hamiltonian. This allows a simultaneous diagonalization of the two parts. The second fact is {\it Griffith's lemma} that we recall for completeness (for a proof see \cite{griffiths} or \cite{DLS}):\\

\noindent{\bf Griffith's Lemma}\cite{griffiths} 
{\it Let $\mu_{n}$ be a sequence of probability measures on $\mathbb R$ such 
that
\begin{equation}f(y)=\lim_{n\to\infty}\frac 1n\ln\int e^{xy}d\mu_{n}(x)\end{equation}
exists for all $y$ in an interval around 0. Define
\begin{equation}a_{\pm}=\lim_{y\to 0^+}\frac{f(\pm y)-f(0)} y.\end{equation}
Then there is a $c<1$ such that, for any $\varepsilon >0$,
\begin{equation}\int_{(a_{-}-\varepsilon)n}^{(a_{+}+\varepsilon)n}d\mu_{n}(x)=1-O(c^n).\end{equation}.}\\

In other words: The scaled measure, $d\mu_{n}(n\xi)$, is 
supported in $a_{-}\leq \xi\leq a_{+}$, with exponential accuracy, as 
$n\to \infty$.

To apply this lemma to the situation under consideration  we take  $x=B$, 
$f(B)=g(0)-g(B)$, $n=|\Lambda|$ and  $ 
d\mu_{n}(n\xi)$ the probability distribution at $B=0$ for the magnetic moment per spin, $\xi=m=|\Lambda|^{-1}M$. Then $a_\pm={} m_\pm$. Moreover, $f$ is a convex function of $B$ so the derivative with respect to  $B$ is monotonously increasing. We can now conclude two things: 1) At points where $f$ is differentiable, i.e., for almost all $B\neq 0$, the probability distribution of the magnetization is a delta function concentrated at ${} m(B)$ and hence ${}{m^2}(B) =\left({} m(B)\right)^2$.
2) The probability distribution for the magnetization at $B=0$ in the thermodynamic limit is concentrated in the interval $[{} m_-,{} m_+]$. Since by convexity of $f$ we have ${} m_+\leq {} m(B)$ for all $B>0$, the statements  1) and 2) together imply ${}{m^2}(0)\leq {}{m^2}(B)$ for $B>0$. Altogether Eq.\  (\ref{grmagn}) is thus established. It is also clear that unless the probability distribution is concentrated at the end points of the interval $[{} m_-,{} m_+]$ the inequality ${}{m^2}(0)\leq {}{m^2}(B)$ is strict. This is exactly what happens for the Heisenberg ferromagnet because as remarked  in \cite{griffiths} the probability distribution is a monotonously decreasing function of $|m|$.

\section{Bosonic Hamiltonians}

We start with the well-known Hamiltonian for bosons in a large box of
volume $V$, expressed in terms of the second-quantized creation\index{creation operator} and
annihilation operators\index{annihilation operator} $\ak, \aks$ satisfying the canonical
commutation relations,
\begin{equation}
H = \sum_\kk k^2 \aks \ak + \frac{1}{2V}\sum_{\kk, \pp, \qq}
\nu(\pp) a_{\kk + \pp}^* a_{\qq - \pp}^* \ak \aq  
\end{equation}
(with $\hbar = 2m =1$). Here, $\nu$ is the Fourier transform of the
two-body potential $v(\x)$.  We assume that there is a bound on the
Fourier coefficients $|\nu(\kk)| \leq \varphi <\infty$.
The case of
hard core potentials\index{hard core interaction} can be taken care of by cutting off the potential at some finite value that is taken to infinity at the end of the calculations.

We shall work in the grand canonical ensemble and hence define $H_\mu=H-\mu N$ where 
$\mu$ is a chemical potential and $N=\sum_{\kk}\aks \ak $ the number operator that generates the gauge symmetry. We also add a gauge symmetry breaking term and define
\begin{equation}H_{\mu,\lambda}= H_\mu+\sqrt{V}(\lambda \aoo
+\lambda^*\aos)\end{equation}
with a complex parameter $\lambda$ that we can take to be a real number without loss of generality. While $H_\mu$ commutes with $N$  this is no longer the case for $H_{\mu,\lambda}$ if $\lambda\neq 0$.

We now come to the $c$-number substitution for $\aoo$ and $\aos$. Let $z$ be a complex number, 
$|z\rangle = \exp\{-|z|^2/2 +z \aos \}\, |0\rangle$ the
coherent state vector in the $\aoo$ Fock space and let $\Pi(z)
=|z\rangle\langle z|$ be the projector onto this vector. There are six
relevant operators containing $\aoo$ in $H_{\mu,\lambda}$, which have the following
expectation
values \cite{klauder} (called {\it lower symbols})\index{lower symbol}
\begin{align}\nonumber
\langle z| \aoo  |z \rangle &= z,
& \!\!\!\langle z| \aoo \aoo |z  \rangle &= z^2, &\!\!\!
\langle z| \aos \aoo |z  \rangle &= |z|^2 \\
\langle z| \aos  |z \rangle &= z^*, & \!\!\!
 \langle z| \aos \aos  |z  \rangle & = z^{*2}, & \!\!\!
\langle z|  \aos \aos \aoo \aoo |z  \rangle &= |z|^4. \nonumber
\end{align}
Each also has an {\it upper symbol}\index{upper symbol}, which is a function of $z$ (call
it $u(z)$ generically) such that an operator $F$ is represented as $F
= \int d^2\!z\, u(z) \Pi(z)$, where $d^2\!z\,\equiv \pi^{-1} dxdy$ with $z=x+iy$. These symbols are
\begin{align}
\aoo &\to z, & \aoo\aoo &\to z^2, &  \aos \aoo &\to |z|^2-1 \nonumber \\
 \aos &\to z^*, & \!\!\!  \aos \aos &\to z^{*2}, &\!\!\!  \aos \aos \aoo
\aoo
 &\to |z|^4 -4|z|^2 +2. \nonumber
\end{align}
We denote by $H_{\mu,\lambda}'(z)$ and $H_{\mu,\lambda}^{\prime\prime}(z)$ the operators obtained from $H_{\mu,\lambda}$ by replacing the polynomials in $\aoo$ und $\aos$ by their lower and upper symbols respectively. These operators act on the Fock-space\index{Fock space} of all the modes other than the $\aoo$ mode. One has 
\begin{equation}H^{\prime
  \prime}_\mu(z) = H^\prime_\mu(z) + \delta_\mu(z)\end{equation} with
\begin{equation}\label{delta}
 \delta_\mu(z) = \mu+ \frac{1}{2V}\Big[ (-4|z|^2 +2) \nu({\bf 0})
- \sum_{\kk \neq {\bf 0}}
a_{\kk}^*a_{\kk}^{\phantom *}\big(2 \nu({\bf 0}) + \nu(\kk)+
\nu(-\kk)\big) \Big] .
\end{equation}

The partition functions\index{partition function} corresponding to these operators are defined  by
\begin{align}\label{partition}
e^{\beta V p(\mu,\lambda)} \equiv \Xi({\mu,\lambda}) &= \Tr_\hi \exp[ -\beta H_{\mu,\lambda}] \\
e^{\beta V p'({\mu,\lambda})} \equiv\Xi'({\mu,\lambda}) &= \int d^2\!z\, \Tr_\hip
\exp[ -\beta H_{\mu,\lambda}^{\prime}(z)]\\e^{\beta V p^{\prime\prime}({\mu,\lambda})} \equiv\Xi^{\prime\prime}({\mu,\lambda}) &= \int d^2\!z\, \Tr_\hip
\exp[ -\beta H_{\mu,\lambda}^{\prime\prime}(z)]
\label{partitionint}
\end{align}
where $\hi$ is the full Hilbert (Fock) space and $\hip$ is the Fock space
without
the $\aoo$ mode. The
functions  $p(\mu,\lambda)$, $p'(\mu,\lambda)$ and $p^{\prime\prime}({\mu,\lambda})$  are the corresponding finite volume pressures and we shall soon see that in the  thermodynamic limit they all coincide. This follows from the inequalities 
\begin{equation}\label{correx0}
\Xi'(\mu,\lambda) \leq \Xi(\mu,\lambda) \leq \Xi''(\mu,\lambda)\leq \Xi'(\mu + 2\varphi/V,\lambda) e^{\beta (|\mu|+\varphi/V)}
\end{equation}
which we now prove. The first step is
\begin{equation}\label{clowerbound}
\Xi(\mu,\lambda) \geq \Xi'(\mu,\lambda),
\end{equation}
which is a consequence of the following two facts: The completeness
property of coherent states,
$ \int d^2\!z\, \Pi (z) = {\rm Identity}$, and
\begin{equation} \label{jensen}
\langle z\otimes \phi | e^{-\beta H_{\mu,\lambda}} | z\otimes \phi \rangle \geq
e^{-\beta \langle z\otimes \phi | H_{\mu,\lambda} |z \otimes \phi \rangle } =
 e^ {-\beta \langle \phi |H_{\mu,\lambda}^\prime ( z) |\phi \rangle },
\end{equation}
where $\phi$ is any normalized vector in $\hip$. This
 is Jensen's inequality for the expectation value of a
convex function (like the exponential function) of an operator.
To prove (\ref{clowerbound}) we take $\phi$ in (\ref{jensen}) to be one
of the normalized eigenvectors of $H_{\mu,\lambda}^\prime (z)$, in which case
$\exp\{\langle \phi |-\beta H_{\mu,\lambda}^\prime ( z) |\phi \rangle\} =
\langle \phi | \exp\{-\beta H_{\mu,\lambda}^\prime ( z) \}|\phi \rangle $.  We
then sum over all such eigenvectors (for a fixed $z$) and integrate
over $z$.  The left side is then $\Xi(\mu,\lambda)$, while the right side is
$\Xi'(\mu,\lambda)$.

The second inequality is the Berezin-Lieb inequality \cite{BL,Li,simon} 
\begin{equation}\label{upper}
\Xi(\mu,\lambda) \leq \Xi''(\mu,\lambda).\end{equation}
 Its
proof is the following. Let $|\Phi_j \rangle \in \hi$ denote the
complete set of normalized eigenfunctions of $H_{\mu,\lambda}$. The partial
inner product $|\Psi_j(z)\rangle = \langle z| \Phi_j\rangle $ is a
vector in $\hip$ whose square norm, given by $c_j(z) = \langle \Psi_j (z) |
\Psi_j(z) \rangle_\hip$, satisfies $\int d^2\!z\, c_j(z) =1$. By using the
upper symbols, we can write 
$$\langle \Phi_j | H_{\mu,\lambda} | \Phi_j\rangle =\int
d^2\!z\, \langle \Psi_j (z) | H_{\mu,\lambda}^{\prime \prime} (z) | \Psi_j
(z)\rangle = \int
d^2\!z\,\langle \Psi_j' (z) | H_{\mu,\lambda}^{\prime \prime} (z)|
\Psi_j'(z)\rangle c_j(z), $$ where $|\Psi_j'(z) \rangle$ is the
normalized vector $c_j(z)^{-1/2} \Psi_j(z)$.  To compute the trace, we
can exponentiate this to write $\Xi(\mu,\lambda)$ as
\begin{equation}\nonumber
 \sum_j \exp\left\{-\beta \int d^2\!z\, c_j(z)
\langle \Psi_j' (z) | H_{\mu,\lambda}^{\prime \prime} (z)|
\Psi_j'(z)\rangle  \right\}.
\end{equation}
Using Jensen's inequality twice, once for functions and once for
expectations as in (\ref{jensen}), $\Xi(\mu,\lambda)$ is less than
\begin{align} \nonumber
 &\sum_j \int d^2\!z\, c_j(z) \exp\left\{
\langle \Psi_j' (z) | -\beta H_{\mu,\lambda}^{\prime \prime} (z) |
\Psi_j'(z)\rangle \right\} \\ &\leq \sum_j \int d^2\!z\, c_j(z)
\langle \Psi_j' (z) | \exp\left\{-\beta H_{\mu,\lambda}^{\prime \prime} (z) \right\}
|
\Psi_j'(z)\rangle .\nonumber
\end{align}
Since $\Tr\, \Pi(z) = 1$, the last
expression can be rewritten as
\begin{equation}\nonumber
\int d^2\!z\, \sum_j \langle \Phi_j |  \Pi(z) \otimes
\exp\left\{-\beta H_{\mu,\lambda}^{\prime \prime}(z)
\right\}| \Phi_j \rangle = \Xi''(\mu,\lambda)
\end{equation}
and (\ref{upper}) is proved.

For the last inequality in (\ref{correx0})  we have to bound $\delta_\mu(z)$ in (\ref{delta}). This is
easily done in terms of the total number operator whose lower symbol
is $N^{\prime}(z) = |z|^2 + \sum_{\kk \neq {\bf 0}}\aks \ak$. In
terms of the bound $\varphi$ on $\nu(\pp)$
\begin{equation}\label{bounddelta}
|\delta_\mu(z)|  \leq 2\varphi (N'(z)+\half )/V +|\mu| \ .
\end{equation}
Consequently, $ \Xi''(\mu,\lambda)$ and
$\Xi'(\mu,\lambda) $ are related by the inequality
\begin{equation}
\Xi''(\mu,\lambda) \leq \Xi'(\mu + 2\varphi/V,\lambda) e^{\beta (|\mu|+\varphi/V)}.
\label{correx1}
\end{equation}

Closely related to this point is the question of relating $\Xi(\mu,\lambda)$
to the maximum value of the integrand in (\ref{partitionint}), which
is $\max_z \Tr_\hip \exp[ -\beta H_{\mu,\lambda}^\prime(z)] \equiv \exp(\beta V
  p^{\max})$.  This latter quantity is often used in discussions of
the $z$ substitution problem, e.g., in refs. \cite{gin,zag}.  One
direction is not hard.  It is the inequality (used in ref.\ \cite{gin})
\begin{equation}\label{junk}
\Xi(\mu,\lambda) \geq \max_z \Tr_\hip  \exp[ -\beta H_{\mu,\lambda}^\prime(z)],
\end{equation}
and the proof is the same as the proof of (\ref{clowerbound}), except
that this time we replace the completeness relation for the coherent
states by the simple inequality ${\rm Identity} \geq \Pi(z)$ for
any fixed number $z$.

For the other direction, split
the integral defining $\Xi''(\mu,\lambda)$ into a part where $|z|^2 <  \xi$ and
$|z|^2\geq \xi$. Thus,
\begin{equation}\label{cheb}
  \Xi''(\mu,\lambda) \leq  \xi \max_z \Tr_\hip  \exp[ -\beta
H_{\mu,\lambda}^{\prime\prime}(z)]
+  \frac 1{\xi} \int_{|z|^2\geq \xi} d^2\!z\,
  |z|^2 \, \Tr_\hip \exp[ -\beta H_{\mu,\lambda}^{\prime\prime}(z)].
\end{equation}
Dropping the condition $|z|^2\geq \xi$ in the last integral and using
$|z|^2\leq N'(z)=N''(z)+1$, we see that the last term on the right side of  (\ref{cheb}) is
bounded above by $\xi^{-1} \Xi''(\mu,\lambda) [V\rho''(\mu,\lambda)+1]$, where
$\rho''(\mu,\lambda)$ denotes the density in the $H_{\mu,\lambda}''$ ensemble. Optimizing
over $\xi$  leads to
\begin{equation}\label{morejunk}
\Xi''(\mu,\lambda) \leq 2 [V \rho''(\mu,\lambda)+1]  \, \max_{z} \Tr_\hip  \exp[ -\beta
H_{\mu,\lambda}^{\prime\prime}(z)].
\end{equation}
Note that $\rho''(\mu,\lambda)$ is order one, since $p''(\mu,\lambda)$ and $p(\mu,\lambda)$
agree in the thermodynamic limit (and are convex in $\mu$), and we assume that the density
in the original
ensemble is finite. By (\ref{bounddelta}), $H_{\mu,\lambda}^{\prime\prime}\geq
H_{\mu+2\varphi/V}^\prime-|\mu|-\varphi/V$, and  it follows from
(\ref{upper}), (\ref{morejunk}) and~(\ref{junk}) that $p^{\max}$
agrees with the true pressure $p$ in the thermodynamic limit. Their
difference, in fact, is at most $O(\ln V /V)$. This is the result obtained
by
Ginibre in \cite{gin} by more complicated arguments, under the
assumption of super-stability\index{superstability} of the interaction, and without the
explicit error estimates obtained here.

To summarize the situation so far, we have four expressions for the
grand-canonical pressure and they are all equal in the thermodynamic limit
\begin{equation}\label{equality}
p(\mu,\lambda) = p'( \mu,\lambda) = p''(\mu,\lambda) =  p^{\max}(\mu,\lambda)
\end{equation}
when $(\mu,\lambda)$ is not a point at which the density can be infinite.\\

The expectation values $\langle \aos\aoo\rangle_{\mu,\lambda}$ and $\langle
\aoo\rangle_{\mu,\lambda}$ are obtained by integrating $(|z|^2-1)$ and
$z$, respectively, with the weight $W_{\mu,\lambda}(z)$, given by 
$$W_{\mu,\lambda}(z)\equiv
\Xi(\mu,\lambda)^{-1} \Tr_{\hip} \langle z| \exp\{-\beta
H_{\mu,\lambda}\}| z\rangle.$$
We will show that for almost every $\lambda$,
the density $W_{\mu,\lambda}(\zeta\sqrt V)$ converges in the thermodynamic limit  to a $\delta$-function
at
the point $\zeta_{\rm max}=\lim_{V\to\infty} z_{\rm max}/\sqrt V$, where
$z_{\rm
  max}$ maximizes the partition function $\Tr_{\hip} \exp\{ -\beta
H'_{\mu,\lambda}(z)\}$. 

That is,
\begin{equation}\label{19}
 V^{-1} \langle \aos\aoo\rangle_{\mu,\lambda}=
 V^{-1} |\langle \aoo\rangle_{\mu,\lambda}|^2
=V^{-1} |z_{\max} |^2
\end{equation}
in the thermodynamic limit. This holds for those $\lambda$ where the pressure in the thermodynamic limit
is differentiable; since $p(\mu,\lambda)$ is convex (upwards) in
$\lambda$ this is true almost everywhere. The right and left
derivatives exist for every $\lambda$ and hence the quasi average
$\lim_{\lambda \to 0+} \lim_{V\to\infty} V^{-1} |\langle
\aoo\rangle_{\mu,\lambda}|^2$ exists.

If we could replace $W_{\mu,\lambda}(z)$ by $W_{\mu,0}(z)e^{-\beta
  \lambda\sqrt{V}(z+z^*)}$, the convergence to a $\delta$-function  would follow from Griffiths'
argument in the same way as for the magnetic model in the previous section.  Because $[H,\aoo]\neq 0$, $W_{\mu,\lambda}$ is
not of this product form.  However, the weight for
$\Xi''(\mu,\lambda)$, which is
$$W''_{\mu,\lambda}(z)\equiv
\Xi''(\mu,\lambda)^{-1} \Tr_{\hip}\exp\{-\beta
H^{\prime\prime}_{\mu,\lambda}(z)\},$$ does have the right form. In the
following we shall show that the two weights are equal apart from
negligible errors.

Equality (\ref{equality}) holds also for all $\lambda$, i.e.,
$p(\mu,\lambda) = p''( \mu,\lambda) = p^{\max}(\mu,\lambda)$ in the
thermodynamic limit.  In fact, since the upper and lower symbols agree for $\aoo$ and
$\aos$, the error estimates above remain unchanged.  (Note that since
$\sqrt{V}|\aoo+\aos|\leq \delta (N+\half) + V/\delta$ for any $\delta>0$,
$p(\mu,\lambda)$ is finite for all $\lambda$ if it is finite for
$\lambda=0$ in a small interval around $\mu$.) At any point of
differentiability with respect to $\lambda$, Griffiths' Lemma
\cite{griffiths} (see the previous section), applied to the partition
function $\Xi''(\mu,\lambda)$, implies that $W''_{\mu,\lambda}(\zeta \sqrt
V)$ converges to a $\delta$-function at some point $\widehat \zeta$ on the
real axis as $V\to\infty$. (The original Griffiths argument can easily
be extended to two variables, as we have here. Because of radial
symmetry, the derivative of the pressure with respect to ${\rm Im\,
}\lambda$ is zero at any non-zero real $\lambda$.)  Moreover, by
comparing the derivatives of $p''$ and $p^{\max}$ we see that
$\widehat \zeta= \lim_{V\to\infty} z_{\rm max}/\sqrt{V}$, since $z_{\rm
  max}/\sqrt{V}$ is contained in the interval between the left and
right derivatives of $p^{\max}(\mu,\lambda)$ with respect to
$\lambda$.

We shall now show that the same is true for $W_{\mu,\lambda}$. To this
end, we add another term to the Hamiltonian, namely $\eps F\equiv \eps V
\int
d^2\!z\, \Pi(z) f(zV^{-1/2})$, with $\eps$ and $f$ real. If $f(\zeta)$ is
a
nice function of two real variables with bounded second
derivatives, it is then easy to see that the upper and lower symbols of
$F$ differ only by a term of order $1$. Namely, for some $C>0$
independent of $z_0$ and $V$,
\begin{equation}\nonumber
\left| V \int d^2\!z\, |\langle z|z_0\rangle |^2
\left( f(zV^{-1/2})- f(z_0 V^{-1/2})\right) \right| \leq C .
\end{equation}
Hence, in particular,
$p(\mu,\lambda,\eps)=p''(\mu,\lambda,\eps)$ in the TL. Moreover, if
$f(\zeta)=0$ for $|\zeta-\widehat \zeta|\leq \delta$, then the pressure is
independent of $\eps$ for $|\eps|$ small enough (depending only on
$\delta$). This can be seen as follows. We have
\begin{equation}\label{c21}
p''(\mu,\lambda,\eps) - p''(\mu,\lambda,0) = \frac 1{\beta V} \ln
\left\langle
e^{-\beta\eps V f(zV^{-1/2})} \right\rangle,
\end{equation}
where the last expectation is in the $H_\mu''$ ensemble at $\eps =0$.  The
corresponding distribution is exponentially localized at
$z/\sqrt{V}=\widehat
\zeta$ by Griffiths' Lemma, and therefore the right side of (\ref{c21})
goes to zero in the thermodynamic limit for small enough $\eps$.  In particular, the
$\eps$-derivative of the  thermodynamic limit pressure at $\eps=0$ is zero.  By
convexity in $\eps$, this implies that the derivative of $p$ at finite
volume, given by $V^{-1}\langle F\rangle_{\mu,\lambda} = \int
d^2\!z\,f(zV^{-1/2}) W_{\mu,\lambda}(z)$, goes to zero in the thermodynamic limit.
Since $f$ was arbitrary, $V\int_{|\zeta-\widehat \zeta|\geq \delta}
d^2\!\zeta\,
W_{\mu,\lambda}(\zeta\sqrt V)\to 0$ as $V\to \infty$. This holds for all
$\delta>0$, and therefore proves the statement.

Our method also applies to the case when the pressure is not
differentiable in $\lambda$ (which is the case at $\lambda=0$ in the
presence of BEC). In this case, the resulting weights
$W_{\mu,\lambda}$ and $W''_{\mu,\lambda}$ need not be
$\delta$-functions, but as in the argument for (\ref{grmagn}) in the previous section Griffiths' method implies
that they are, for $\lambda\neq 0$, supported on the real axis between
the right and left derivative of $p$ and, for $\lambda=0$, on a disc
(due to the gauge symmetry)\index{gauge symmetry} with radius determined by the right
derivative at $\lambda=0$.  Convexity of the pressure as a function of
$\lambda$ thus  implies that in the thermodynamic limit  the supports of the weights $W_{\mu,\lambda}$
and $W''_{\mu,\lambda}$ for $\lambda\neq 0$ lie outside of this disc.
Hence $\langle \no\rangle_\lambda$ is monotone increasing in $\lambda$
in the thermodynamic limit. In combination with (\ref{19}) this implies in particular that
\begin{equation}\label{roep}
\lim_{V\to\infty} \frac 1V \langle
\aos\aoo\rangle_{\mu,\lambda=0}\leq \lim_{\lambda\to 0} \lim_{V\to \infty}
\frac 1V \langle \aos\aoo\rangle_{\mu,\lambda} =\lim_{\lambda\to 0} \lim_{V\to \infty}
\frac 1V |\langle \aoo\rangle_{\mu,\lambda}|^2 .
\end{equation}
Hence (\ref{mainstatement}) is established.

We note that by Eq. (\ref{roep}) spontaneous symmetry
breaking\index{spontaneous symmetry breaking} (in the sense that the
right side of (\ref{roep}) is not zero) always takes place whenever
there is BEC is the usual sense, i.e., without explicit gauge breaking
(meaning that the left side of (\ref{roep}) is non-zero). Note, however, that a non-vanishing of the right side of (\ref{roep})
does not {\it a priori} imply a non-vanishing of the left side. I.e.,
it is {\it a priori} possible that BEC only shows up after introducing
an explicit gauge-breaking term to the Hamiltonian. While it is expected on
physical grounds that positivity of the right side of  (\ref{roep}) implies positivity of the left side, a rigorous proof is lacking, so far. In the
example of the Heisenberg magnet discussed in the previous section, equality in
(\ref{roep}) does not generally hold, but still both sides are
non-vanishing in the same parameter regime.

To illustrate what could  arise mathematically, in principle, consider a weight function of the form
\begin{equation}\label{junk5}
W''_{\mu,\lambda=0} (\sqrt V \zeta) \equiv  w_V(\zeta) = \left\{ \begin{array}{cl}  V^2 - V + 1/V & {\rm for\ } |\zeta|\leq 1/V \\ 1/V & {\rm for\ } 1/V \leq |\zeta|\leq 1 \\ 0 & {\rm for\ } |z|>1\,. \end{array} \right.
\end{equation}
This distribution converges for $V\to \infty$ to a $\delta$-function at $\zeta=0$, and
consequently there is no BEC at $\lambda=0$. On the other hand, it is
easy to see that the weight function $w_V(\zeta)e^{-\beta \lambda V \zeta}$
(with an appropriate normalization factor) converges, for any
$\lambda>0$, to a $\delta$-function at $\zeta=-1$ as $V\to\infty$, and hence there is
spontaneous symmetry breaking. An  open problem for the mathematician is to prove that examples like (\ref{junk5}) do not occur in realistic bosonic systems.

\end{document}